\documentclass[reprint, 
  superscriptaddress,
  amsmath,
  amssymb,
  aps, 
  showkeys,
  ]{revtex4-2}

\usepackage{graphicx}
\usepackage{dcolumn}
\usepackage{bm}
\usepackage{makecell}   
\usepackage{multirow}
\usepackage{xcolor}
\usepackage[hidelinks]{hyperref}

\begin{document}

\title{Modeling partially-ionized dense plasma using wavepacket molecular dynamics}

\author{Daniel Plummer}
\email{daniel.plummer@physics.ox.ac.uk}
\affiliation{Department of Physics, University of Oxford, UK}

\author{Pontus Svensson}
\affiliation{Department of Physics, University of Oxford, UK}
\affiliation{Center for Advanced Systems Understand (CASUS), D-02826 G\"orlitz, Germany}
\affiliation{Helmholtz-Zentrum Dresden-Rossendorf (HZDR), D-01328 Dresden, Germany}

\author{Wiktor Jasniak}
\affiliation{Department of Physics, University of Oxford, UK}

\author{Patrick Hollebon}
\affiliation{AWE, Aldermaston, Reading, Berkshire RG7 4PR, UK}

\author{Sam M. Vinko}
\affiliation{Department of Physics, University of Oxford, UK}

\author{Gianluca  Gregori}
\affiliation{Department of Physics, University of Oxford, UK}

\date{\today}


\begin{abstract}
  We develop a wavepacket molecular dynamics framework for modeling the structural properties of partially-ionized dense plasmas, based on a chemical model that explicitly includes bound state wavefunctions. Using hydrogen as a representative system, we compute self-consistent charge state distributions through free energy minimization, following the approach of Plummer \emph{et al.} [Phys. Rev. E \textbf{111}, 015204 (2025)]. This enables a direct comparison of static equilibrium properties with path integral Monte Carlo data, facilitating an evaluation of the model’s underlying approximations and its ability to capture the complex interplay between ionization and structure in dense plasma environments.
\end{abstract}

\maketitle

\section{Introduction} \label{sec:Introduction} 
Warm dense matter (WDM) refers to a state of matter that occupies the intermediate regime between condensed matter and hot plasma. It is characterized by near-solid densities and temperatures ranging from several thousand up to hundreds of thousands of Kelvin and encompasses systems such as dense partially-ionized plasmas. These conditions provide an intriguing but formidable challenge for model development, given that an interplay of partial ionization, strong inter-particle correlations, and electron degeneracy influence physical properties in this regime~\cite{Bonitz2020,bonitz2024pub}.

As an atomistic technique, semi-classical molecular dynamics bridges the gap between first-principles calculations and hydrodynamic simulations, which are needed for large-scale, long-time fluid evolution beyond particle-resolved methods. Additionally, they may provide access to properties out of reach for first-principles methods, including path integral Monte Carlo and  density functional theory approaches~\cite{bonitz2024pub}. However, the rich variety of physical processes present in dense plasma often render semi-classical models inaccurate, and they must be benchmarked to assess their underlying validity, with empirical adjustments made when necessary. One such class of models is the wavepacket molecular dynamics (WPMD) method  which provides a semi-classical description of an electron-ion plasma as an interacting gas of parameterized wavepackets (the electrons) and point charges (the ions). In this approach, the system is evolved based on equations of motion derived from a time-dependent variational principle. This principle is motivated by the Schrödinger equation, and captures quantum effects approximately by restricting the dynamics to a chosen variational ansatz. The WPMD method has been applied to a diverse range of phenomena such as computing the equation of state~\cite{Su2007, Lavrinenko2021,Jakob2007}, prediction of x-ray Thomson scattering spectra and associated dynamic structure~\cite{Svensson2024pub, Zwicknagel2006, Davis2020}, and investigating both ion and electronic transport coefficients~\cite{Klakow1994b, Svensson2025,Morozov2009,Angermeier2023,Yao2021}. For additional references see the review by Grabowski~\cite{Grabowski2014}. 

This work is concerned with an extension of WPMD to a partially-ionized plasma and the resulting impact on both the structural properties and non-ideal ionization effects. To this end, an explicit model for bound electrons is presented, inspired by Ref.~\cite{Ebeling1997}. The self-consistent equilibrium charge state distribution of the model is then calculated based on the free energy minimization framework developed in Ref.~\cite{Plummer2025}. Specifically, hydrogen is chosen as a test case due to its simple atomic structure and the availability of high-fidelity reference data for structural properties in the partially-ionized regime~\cite{Dornheim2024}. Furthermore, through the lens of this investigation the dependence of the resulting properties on the electron extent, adjusted through a harmonic confining potential, and the model-dependent nature of ionization are also brought into focus. The manuscript is structured as follows: In Sec. \ref{sec:boundStateModel} bound states are introduced into WPMD, including the treatment of exchange effects through appropriate Pauli potentials. Following this the numerical implementation is elaborated in Sec. \ref{sec:numerical}. Sec. \ref{sec:confining} details the role of the confining potential which acts to regularize the model. In Sec. \ref{sec:ionization} the ionization calculations are presented accompanied by the resulting impact on structural properties. Finally, concluding remarks are given in Sec. \ref{sec:conclusion}.

\begin{widetext}
\section{\label{sec:boundStateModel} Bound State Model For wavepacket molecular dynamics}

In this section the underlying theoretical description of wavepacket models is reviewed, before the appropriate modifications to include bound electrons are introduced. 

\subsection{Ehrenfest molecular dynamics}
We start at the level of Ehrenfest molecular dynamics~\cite{Hutter2009}, by considering the nonadiabatic time evolution of a charge-neutral system of \(N\) classical protons coupled to \(N\) quantum-mechanical electrons. By assuming the classical particles to be protons, we specialize to the case of hydrogen, although the generalization to heavier elements is straightforward. The proton phase space is therefore parameterized by the spatial coordinates and momenta \(\{\boldsymbol{R}_I(t), \boldsymbol{P}_I(t)\}\) where the index \(I = 1 \dotsc N\) labels each proton. Meanwhile the time-dependent electronic state is the \(N\)-electron many-body wavefunction \(|\Psi_e\rangle\), a vector in the corresponding Hilbert space. The classical Hamiltonian that determines both the total energy and the evolution of the protons is
\begin{equation} \label{eq:Ehrenfest_classical_hamiltonian}
  \mathcal{H}(\{\boldsymbol{R}_I(t)\}, \{\boldsymbol{P}_I(t)\};\Psi_e(t)) = \sum_I \frac{\boldsymbol{P}_I^2}{2M_p} + \frac{e^2}{4\pi \varepsilon_0 }\sum_{I<J} \frac{1}{|\boldsymbol{R}_I - \boldsymbol{R}_J|} + \langle \Psi_e | \hat{H}_e(\{\boldsymbol{R}_I\}) |\Psi_e\rangle
\end{equation}
where \(M_p\) is the proton mass, and it is clear that the ions evolve in the time-dependent mean-field potential of the electrons and under their own Coulomb interactions. Furthermore, \(\hbar\) is the reduced Planck's constant, \(\epsilon_0\) is the permittivity of free space, \(e\) is the electron charge and \(m_e\) is the electron mass.  For a non-relativistic Coulomb system, the electronic Hamiltonian is:
  \begin{equation} \label{eq:electronic_hamiltonian}
\hat{H}_e(\{\boldsymbol{R}_I\}) = 
- \sum_i \frac{\hbar^2}{2m_e} \nabla_i^2 
+ \frac{e^2}{4\pi \varepsilon_0 }\sum_{i<j} \frac{1}{|\hat{\boldsymbol{x}}_i - \hat{\boldsymbol{x}}_j|} 
- \frac{e^2}{4\pi \varepsilon_0 }\sum_{I,i} \frac{1}{|\boldsymbol{R}_I - \hat{\boldsymbol{x}}_i|} 
\end{equation}

\end{widetext} 
where the lower case indices label electrons, \(\nabla_i\) is the gradient operator with respect to the coordinates of the \(i\)-th electron and \(\hat{\boldsymbol{x}}_i\) is the position operator. Meanwhile, the electron dynamics are governed by the time-dependent many-electron Schr\"odinger equation,
\begin{equation} \label{eq:schrodinger_electrons} 
  i \hbar \frac{\partial \Psi_e }{\partial t} = \hat{H}_e \Psi_e.
\end{equation}
Ehrenfest molecular dynamics is used in its exact form within quantum chemistry, where small numbers of degrees of freedom may be studied on short timescales~\cite{Hutter2009}. To access the dynamics of larger systems over longer timescales, the required particle numbers mean that further approximations must be made. 

\subsection{Mixed bound-free functional form}
In the WPMD method~\cite{Klakow1994b,Feldmeier2000}, the electronic wavefunction is modeled as a parameterized function, in other words the following replacement is performed:
\begin{equation}
  |\Psi_e \rangle \rightarrow |Q(\{{q_\mu}\})\rangle
\end{equation}
where \(\{{q_\mu}\}\) are a set of state parameters, henceforth constrained to be real. The equations of motion can be derived from a generalization of Eq. \eqref{eq:schrodinger_electrons}, derived from the following action functional~\cite{Feldmeier2000}:
\begin{equation} \label{eq:TDVP}
  S = \int dt \mathcal{L}(\{q_\mu\}, \{\dot{q}_\mu\}) = \int dt \langle Q | i \hbar \frac{d}{d t} - \hat{H}_e | Q \rangle,
\end{equation}
where \(\mathcal{L}\) is the Lagrangian and \(|Q\rangle\) is the parameterized many-electron state with the \(q_\mu\)-dependence now implicit. A time-dependent variational principle may now be defined by requiring the action to be stationary under variations, \(\delta S = 0\), which yields the Euler-Lagrange equations,
\begin{equation} \label{eq:euler_lagrange} \frac{\partial \mathcal{L}}{\partial q_\mu} - \frac{d}{dt}\frac{\partial \mathcal{L}}{\partial \dot{q_\mu}} = 0,\end{equation} 
and reduces to true Schrödinger evolution when the state is unrestricted~\cite{Feldmeier2000}. This variational principle is also employed in real-time time-dependent density function theory calculations~\cite{Hutter2009, Runge1984, Kononov2022}. The electron dynamics are hence constrained to a submanifold of the full Hilbert space, specified by the choice of \(|Q\left(\{q_\mu\}\right)\rangle\). All further approximations depend on how the parameters are chosen, and for plasmas the state is typically a Hartree product of localised Gaussian wavefunction~\cite{Klakow1994b,Svensson2023} -- inspired by the point particle description of classical plasma -- although antisymmetrized forms have been considered for lower temperatures~\cite{Jakob2007}.  In this work we follow Ref.~\cite{Ebeling1997} and employ a mixed Ansatz of \(N_{\text{n}}\) bound electrons with single-particle orbitals \(|b_k\rangle\) and \(N_{\text{e}} = N - N_{\text{n}}\) free electrons with single-particle orbitals \(|q_i\rangle\), and choose the constrained many body wavefunction to be of Hartree form,
\begin{equation}
  |Q\rangle = |q_1 \rangle \otimes |q_2 \rangle \otimes \dotsc \otimes |q_{N_{\text{e}}}\rangle \otimes |b_1 \rangle \otimes \dotsc \otimes |b_{N_{\text{n}}}\rangle.
\end{equation}
Imposing this form establishes a chemical picture~\cite{Ebeling2017} and the ionization state must therefore be determined to model equilibrium systems. Here we determine the ionization through the self-consistent method introduced in Ref.~\cite{Plummer2025}. As an alternative, using the schemes outlined in Refs.~\cite{Feldmeier2000,Ebeling1997}, stochastic transitions between bound and free state manifolds could be considered as an extension to this work. Note that limited effects due to the missing antisymmetric structure are accounted for through the use of Pauli potentials, introduced in the following section. Applying this framework to hydrogen plasma,  we pick the bound state wavefunction as a propagating, isolated hydrogen ground state wavefunction which is attached to a specific proton,
\begin{equation}
    \begin{split}
      \langle \boldsymbol{x} | b_{k} \rangle =& \frac{1}{\sqrt{a_0^3 \pi}} \\ & \times\exp\left[-\frac{|\boldsymbol{x} - \boldsymbol{r}_{k}|}{a_0} + \frac{im_e}{\hbar}\boldsymbol{v}_{k}\cdot (\boldsymbol{x} - \boldsymbol{r}_{k})\right].
    \end{split}
\end{equation}
Together the ion and bound electron pair form a neutral particle. Here \(a_0\) is the Bohr radius, and \(\boldsymbol{r}_{k}, \boldsymbol{v}_{k}\) are the variables describing a neutral particle's position and velocity. This ansatz may be motivated by applying a Galilean transformation to a stationary 1s wavefunction. An important note is that the functional form does not introduce any additional quantum degrees of freedom, because the bound electrons are assumed to adiabatically follow the motion of underlying protons labelled by \(k = 1, \dotsc, N_{\text{n}}\). To model free electrons the anisotropic Gaussian state developed in Ref.~\cite{Svensson2023} is employed,
\begin{equation}
  \begin{split}
    \langle \boldsymbol{x} |q_{j}\rangle = & \left((2\pi)^3 \det({\Sigma}_{j})\right)^{-1/4}\\ & \times \exp\left[-\boldsymbol{\xi}^T_{j} \left(\frac{1}{4}{\Sigma}^{-1}_{j} - \frac{i}{\hbar}{\Pi}_{j}\right)\boldsymbol{\xi}_{j} + i \boldsymbol{p}_{j}^T \boldsymbol{\xi}_{j}\right],
  \end{split}
\end{equation} where \(\Sigma_{j}\) and \(\Pi_{j}\) are three-by-three symmetric matrices that parameterize the time-dependent rotation and elongation of the wavepacket and \(\boldsymbol{p}_{j}\) is the expected momentum. The spatial coordinate \(\boldsymbol{\xi}_{j} = \boldsymbol{x} - \boldsymbol{r}_{j}\) is the relative position vector from the expected position of the wavepacket at \(\boldsymbol{r}_{j}\) and the index \(j = 1, \dotsc N_{\text{e}}\) labels free electrons. Finally, the remaining \(N_{\text{i}} = N - N_{\text{n}}\) protons are relabelled according to the imposed chemical picture with positions \(\boldsymbol{r}_{i}\) and momenta \(\boldsymbol{p}_{i}\) where \(i = 1, \dotsc N_{\text{i}}\), and shall be subsequently referred to as ions.
  
\subsection{Equations of motion and total energy}
The resulting expression for the wavepacket Hamiltonian is simply the Ehrenfest Hamiltonian, eq. \eqref{eq:Ehrenfest_classical_hamiltonian}, evaluated for the restricted wavefunction. Explicitly,
\begin{equation}\label{eq:Ehrenfest_evaluated_hamiltonian}
  \mathcal{H} = \mathcal{H}(\{\boldsymbol{r}_{i}\},\{\boldsymbol{p}_{i}\}, \{\boldsymbol{r}_{k}\},\{\boldsymbol{p}_{k}\} ; |Q\rangle),
\end{equation}
where the \(i\) subscript refers to ions, and the \(k\) subscript refers to neutrals. The Hamiltonian is a functional of the restricted wavefunction \(|Q\rangle\) that depends on the neutral and electron degrees of freedom:
\begin{equation}
  |Q\rangle = |Q(\{\boldsymbol{r}_{j}\},\{\boldsymbol{p}_{j}\},\{\Sigma_{j}\},\{\Pi_{j}\}, \{\boldsymbol{r}_{k}\},\{\boldsymbol{p}_{k}\})\rangle,
\end{equation}
where the \(j\) subscript refers to electrons. The numerical method by which the state averages are computed is explained in Sec. \ref{sec:numerical}. The equations of motion that follow from equations \eqref{eq:Ehrenfest_classical_hamiltonian} and \eqref{eq:euler_lagrange} for the position and momentum degrees of freedom are
\begin{equation}
  \frac{d \boldsymbol{r}_\gamma}{dt} = \frac{\partial \mathcal{H}}{\partial \boldsymbol{p}_\gamma} \quad ; \quad \frac{d \boldsymbol{p}_\gamma}{dt} = -\frac{\partial \mathcal{H}}{\partial \boldsymbol{r}_\gamma}
\end{equation}
for the ions (\(\gamma=i\)) and neutrals (\(\gamma=k\)) and the expected positions and momenta of the electrons (\(\gamma=j\)). Meanwhile the internal electronic degrees of freedom obey the following symmetrized Hamiltonian equations of motion~\cite{Svensson2023}:
\begin{equation} \label{eq:internal_eom}
  \begin{split}
  \frac{d}{dt} \Sigma_{j\alpha\beta} =& \frac{1}{2}\left(\frac{\partial \mathcal{H}}{\partial \Pi_{{j}\alpha\beta}} + \frac{\partial \mathcal{H}}{\partial \Pi_{{j}\beta\alpha}}\right), \\
  \frac{d}{dt} \Pi_{j\alpha\beta} =& -\frac{1}{2}\left(\frac{\partial \mathcal{H}}{\partial \Sigma_{{j}\alpha\beta}} + \frac{\partial \mathcal{H}}{\partial \Sigma_{{j}\beta\alpha}}\right),
  \end{split}
\end{equation}
where \(\Sigma_{j,\alpha\beta} = [\Sigma_{j}]_{\alpha\beta}\) is a component of the symmetric width matrix corresponding to the \(j\)-th electron.

\subsection{Pauli Potentials}

We follow an idea based on the pairwise truncation of the antisymmetrization operator~\cite{Klakow1994b,Grabowski2014} which manifests as an effective Pauli potential being added to the classical Hamiltonian. Some works have been dedicated to studying the behaviour of Pauli potentials in ground state~\cite{Su2007, Angermeier2021}, which have resulted in empirical developments through comparison with high fidelity ground-state data. Along similar lines, some authors have considered additional exchange-correlation energy terms in the wavepacket Hamiltonian, Equation~\eqref{eq:Ehrenfest_evaluated_hamiltonian}, at the level of the ground-state local density approximation~\cite{Lavrinenko2019, Lavrinenko2021}, which may provide a better description at degenerate conditions.

In this paper, we wish to study the transition from ionized plasma to atomic gas, where hydrogen bond formation is minimal due to thermal dissociation. Therefore, to partially account for the Fermionic nature of the electrons, we assign each a constant spin and include an additional potential between same-spin electrons. Following previous works we include the dominant pairwise antisymmetric kinetic energy contributions~\cite{Klakow1994a}. Consider the following two-particle triplet state, constructed from two normalized single-particle orbitals \(|\alpha\rangle\) and \(|\beta\rangle\):
\begin{equation}
  |\Psi_2^{A}\rangle = |\alpha\rangle \otimes |\beta\rangle - |\beta\rangle \otimes |\alpha\rangle.
\end{equation}
An additional pair potential is included that includes kinetic energy contributions to a pairwise antisymmetrized state,
\begin{equation} \label{eq:generalPauliPotential}
  \mathcal{V}^P = \frac{\langle \Psi_2^{A} |\hat{K}_e | \Psi_2^{A} \rangle }{\langle \Psi_2^{A} | \Psi_2^{A} \rangle} - \langle \alpha | \otimes  \langle \beta | \hat{K}_e |\alpha\rangle \otimes |\beta\rangle,
\end{equation}
where the product state contribution implicitly present within the model has been subtracted. Note that while the product state is normalized, an additional factor is applied to normalize the antisymmetrized state which is consistent with previous models~\cite{Klakow1994b,Svensson2023}. When summed over all electron pairs, the form of Eq. \eqref{eq:generalPauliPotential} produces an interaction that depends on all free electron degrees of freedom, and hence increases the complexity of the model significantly. The two particle kinetic energy operator takes the standard form
  \begin{equation}
    \hat{K}_e = \frac{1}{2m_e}\left(\hat{\boldsymbol{p}}_1^2 + \hat{\boldsymbol{p}}_2^2\right).
  \end{equation}
All electrons in the model are assigned a constant spin, and same-spin electrons act through this additional potential while opposite-spin interactions are assumed to be purely electrostatic, neglecting higher-order correlation effects~\cite{Angermeier2021}.

\section{Numerical implementation} \label{sec:numerical}

By including bound state orbitals in the many-body wavefunction, this work proposes an extension to the anisotropic wavepacket model developed in Ref.~\cite{Svensson2023}. Furthermore, the existing \textsc{lammps}~\cite{LAMMPS} implementation has been extended to include the additional bound-state Coulomb and Pauli potentials specified. Embedding our model within the larger \textsc{lammps} ecosystem, allows efficient parallelization through domain decomposition. This permits scalable simulations that can handle large numbers of particles~\cite{Svensson2023}. To further streamline numerical performance, an implementation for graphics processing units, which is possible within \textsc{lammps}, could be employed to access longer timescales.

In this section we detail an on-the-fly scheme to compute the additional Coulomb and Pauli terms, allowing for an efficient numerical implementation.

\subsection{Bound state Coulomb interactions}

 All Coulomb interactions are computed by taking state averages of the electronic Hamiltonian, Eq. \eqref{eq:electronic_hamiltonian}, with respect to the different single-particle orbitals. Therefore the pairwise Coulomb interaction between a neutral with index \(k\) and ion with index \(i\) takes the following form:
\begin{equation}
  \begin{split} \label{eq:ion_neutral_coulomb}
  \mathcal{V}_{in}^{C}(r_{ik}) &= \frac{e^2}{4\pi\epsilon_0}\langle b_{k} | \left(\frac{1}{|\boldsymbol{r}_{k} - \boldsymbol{r}_{i}|} -\frac{1}{|\hat{\boldsymbol{x}} - \boldsymbol{r}_{i}|}\right)|b_{k}\rangle \\
  &= \frac{e^2}{4\pi\epsilon_0} \frac{\exp[{-2r_{ik}/a_0}]}{r_{ik}}\left(1 + \frac{r_{ik}}{a_0}\right).
  \end{split}
\end{equation}
Which solely depends on the ion-neutral radial separation denoted by \(r_{i k} = |\boldsymbol{r}_{i} - \boldsymbol{r}_{k}|\). The resulting pair potential is a screened purely repulsive interaction. Similarly, the Coulomb interaction between two neutral particles \(k\) and \(l\) depends on the radial separation \(r_{k l}\),
\begin{widetext}
\begin{equation} \label{eq:neutral_neutral_coulomb}
  \begin{split}
    \mathcal{V}_{nn}^{C}(r_{kl}) &= \frac{e^2}{4\pi\epsilon_0}\langle b_{k}| \otimes \langle b_{l} | \left(\frac{1}{|\hat{\boldsymbol{x}}_1 - \hat{\boldsymbol{x}}_2|} - \frac{1}{|\boldsymbol{r}_{k} - \hat{\boldsymbol{x}}_{2}|}  - \frac{1}{|\boldsymbol{r}_{l} - \hat{\boldsymbol{x}}_{1}|} + \frac{1}{|\boldsymbol{r}_{k} - \boldsymbol{r}_{l}|}\right)|b_{k}\rangle \otimes | b_{l}\rangle \\ &= \frac{e^2}{4\pi\epsilon_0} \frac{\exp[{-2r_{kl}/a_0}]}{r_{kl}}\left(1 + \frac{5}{8}\left(\frac{r_{kl}}{a_0}\right) - \frac{3}{4}\left(\frac{r_{kl}}{a_0}\right)^2 - \frac{1}{6}\left(\frac{r_{kl}}{a_0}\right)^3\right),
  \end{split}
\end{equation}
which is screened due to the presence of two bound electrons and features a negative (attractive) asymptote, allowing for molecule formation at lower temperatures. These analytical formulae are equivalent to those used in Ref.~\cite{Plummer2025} and provide fast and stable energy and force contributions in the implementation. Given that the free-free and ion-free interactions are already present within the model~\cite{Svensson2023}, the final additional Coulomb interaction is the interaction between a free electron and neutral particle. The state average may be written as an integral over the charge density of a free electron with index \(k\) and ion-neutral interaction: 
\begin{equation} 
  \begin{split}\label{eq:electron_neutral}
  V_{en}^{C} &= \frac{e^2}{4\pi\epsilon_0}\langle q_{j}| \otimes \langle b_{k} | \left(\frac{1}{|\hat{\boldsymbol{x}}_1 - \hat{\boldsymbol{x}}_2|}  - \frac{1}{|\hat{\boldsymbol{x}}_{1} - \boldsymbol{r}_{k}|} \right)|q_{j}\rangle \otimes | b_{k}\rangle \\
  &= \int d\boldsymbol{x}_1 |\langle\boldsymbol{x}_1|q_{j}\rangle|^2 \mathcal{V}_{in}^{C}(|\boldsymbol{x}_1 - \boldsymbol{r}_{k}|),
  \end{split} 
\end{equation}
where the kernel \(\mathcal{V}_{in}^{C}\) is defined in Eq. \eqref{eq:ion_neutral_coulomb}. No closed-form solution is readily available for this integral. To proceed, the integration kernel \(\mathcal{V}_{in}^{C}\) is approximated as a sum of Gaussian modes using the scheme outlined in Ref~\cite{Svensson2023},
\begin{equation} \label{eq:ion_neutral_gauss_decomp}
  \mathcal{V}_{in}^{C} \approx \sum_p c_p \exp[-\alpha_p r^2],
\end{equation} which yields a tractable analytical expression to evaluate the free-neutral energy contributions. Further details on the Gaussian decomposition are given in Appendix~\ref{app:gauss_decomp}, while the final expression for the integral over each Gaussian mode is given in Appendix~\ref{app:electron_neutral_coulomb}. This treatment of Coulomb interactions closely mirrors the treatment of the short-range ion-free and free-free Coulomb interactions~\cite{Svensson2023}.

\subsection{Pauli Interactions}

The general Pauli potential, Eq. \eqref{eq:generalPauliPotential}, may be expressed in terms of single-particle matrix elements,
\begin{align} \label{eq:pauli_expanded}
  \mathcal{V}^{\rm{P}} 
  &=  -\frac{1}{1 - |\langle \alpha | \beta\rangle|^2}\bigg(\text{Re}\left[\langle \alpha | \frac{\hat{\boldsymbol{p}}^2}{m_e}|\beta \rangle \langle \beta| \alpha \rangle\right] - \frac{1}{2}\left(\langle \alpha | \frac{\hat{\boldsymbol{p}}^2}{m_e} | \alpha \rangle + \langle \beta | \frac{\hat{\boldsymbol{p}}^2}{m_e} | \beta \rangle\right) |\langle \alpha | \beta\rangle |^2\bigg).
\end{align}
Given that neither closed-form expressions for the kinetic matrix elements nor the overlap are currently available for the free-bound case, a further decomposition of the bound state orbital in terms of isotropic propagating Gaussians \(|g_{k}\rangle\) is applied to perform on-the-fly calculations,
\begin{equation}
  |b_{k}\rangle = \sum_{p} c_p |g_{k,p}\rangle, 
\end{equation}
where 
\begin{equation} \label{eq:propagating_isotropic_gaussian}
  \langle \boldsymbol{x} | g_{k,p} \rangle = \left(\frac{2\alpha_p}{\pi}\right)^{3/4} \exp\left[-\alpha_p (\boldsymbol{x} - \boldsymbol{r}_{k})^2 + i {m}_e \boldsymbol{v}_{k} \cdot (\boldsymbol{x} - \boldsymbol{r}_{k})\right].
\end{equation}
\end{widetext} 
The \(c_p\) and \(\alpha_p\) are fitting parameters determined to best approximate the bound state. In the non-propagating case this is a common technique within self consistent field calculations, and we have used the decomposition provided by Ditchfield \emph{et al.} based on an energy minimization procedure~\cite{Ditchfield1970}. For all simulations, the free-bound calculation was performed using a six mode decomposition, providing an accurate estimate of the kinetic energy. The numerical implementation permits a reduction in the number of modes, balancing computational efficiency against the accuracy of the bound-state representation. Performing the decomposition enables the matrix elements in Eq. \eqref{eq:pauli_expanded} to be evaluated, with explicit formulae available in Appendix \ref{app:kinetic_matrix_elements}. For the bound-bound interactions, it was numerically confirmed that the velocity dependence is negligible for characteristic neutral thermal velocities at temperatures where partial ionization is expected. Neglecting the velocity dependence permits known formulae for the stationary 1s wavefunction to be used~\cite{alma990165672020107026}, resulting in the following neutral-neutral pair potential,
\begin{equation}
  V^{P}_{nn}(\tilde{r}) = \frac{2\hbar^2}{3 m_e a_0^2}\frac{e^{-2\tilde{r}} \left( \tilde{r}^2 + \tilde{r}^3 + \frac{1}{3} \tilde{r}^4 \right)}{1 - e^{-2\tilde{r}} \left( 1 + 2\tilde{r} + \frac{5}{3} \tilde{r}^2 + \frac{2}{3} \tilde{r}^3 + \frac{1}{9} \tilde{r}^4 \right)}
\end{equation}
which acts between neutrals with same-spin bound electrons. Here \(\tilde{r}=r_{k l}/a_0\) is the radial distance between two neutral particles labelled by \(k\) and \(l\), measured in Bohr radii.

\subsection{Confining potential}

When applying WPMD to systems where ionization effects are present, electrons tend to spread indefinitely thereby diminishing their interactions. To remedy this, a confining potential is often used to empirically adjust the extent of the electrons and regularize the model~\cite{Svensson2023,Svensson2024pub,Grabowski2014,Su2007}. In fact, even in the noninteracting case, the partition function associated to the classical Hamiltonian will diverge when the electrons are unconfined~\cite{Ebeling2006}. In this work a harmonic potential is added to each free electron through the electronic Hamiltonian,
\begin{equation}\label{eq:confining_potential_operator}
  \hat{V}_{\Sigma} = \frac{A}{2} (\hat{\boldsymbol{x}} - \langle\hat{\boldsymbol{x}}\rangle)^2,
\end{equation}
where \(A\) is an empirical parameter, the effect of which is investigated in the following section. Note that the issue of unrestricted spreading does not arise in Refs~\cite{Lavrinenko2016,Lavrinenko2021} when a confined (non-periodic) system is considered. To approximate a bulk plasma, we consider periodic boundary conditions here and hence use the standard confining potential, treating $A$ as an empirical parameter.

Evaluating the state average in Eq. \eqref{eq:confining_potential_operator} for each free electron gives the following contribution to the classical Hamiltonian,
\begin{equation}
  \mathcal{V}_{\Sigma} = \sum_{j=1}^{N_{\text{e}}} \frac{A}{2}\text{Tr}\left(\Sigma_{j}\right),
\end{equation}
which is a special case of the potential already used in Ref.~\cite{Svensson2023}. To estimate a relation between the spatial extent of each free electron and the confinement strength \(A\), the shape kinetic energy for an isotropic wavepacket may be equated with its confining potential. This leads to
\begin{equation}\label{eq:iso_mode}
  A = \frac{\hbar^2}{4m_e\sigma_0^4},
\end{equation}
where \(\sigma_0\) is a representative unbound confinement width characterising the electrons under a given confining strength which shall henceforth be referred to as the \emph{confinement parameter}. A similar characterization of the confining potential has been used previously in Ref.~\cite{Knaup1999}. Eq.~\eqref{eq:iso_mode} corresponds to the modal width of an ensemble of isotropic wavepackets. The associated distributions of isotropic and anisotropic widths are discussed in Ref.~\cite{Plummer2026_statistics_arxiv}.

After presenting general remarks on the molecular dynamics simulations, the implications of confinement for the resulting structural properties are discussed in the following section.

\section{The role of the confining potential} \label{sec:confining}

\subsection{Simulation Details}

To assess the efficacy of the models, we compare against path integral Monte Carlo (PIMC) data from Ref.~\cite{Dornheim2024}. While these data provide a valuable benchmark, they remain affected by finite-size errors arising from the restricted system sizes that are computationally accessible within the PIMC method. These data are available for the structural properties of hydrogen at two conditions, both corresponding to a degeneracy parameter \(\theta = k_B T / E_F\) where \(E_F\) is the Fermi energy, of unity. The low temperature condition corresponds to a Br\"uckner density parameter of \(r_s = 3.23\) and temperature of \(T = 55700 \, \text{K}\).  Here
\begin{equation}
  r_s = \left( \frac{3}{4\pi n} \right)^{1/3} \frac{1}{a_0}
\end{equation} is the Wigner-Seitz radius in units of Bohr radii and the total electron (or proton) number density \(n = N/V\) is the ratio of the total number of electrons (or protons) to the volume of the system. At this condition, hydrogen is generally expected to be a partially-ionized atomic fluid~\cite{Filinov2023,Demyanov2025,Dornheim2024,Filinov2004} and therefore exhibits strong electron-ion coupling. The high-temperature condition corresponds to a density parameter of \(r_s = 2\) and is expected to be fully ionized. At this condition the electron-ion coupling is weak~\cite{Moldabekov2022} and bound state formation is minimal due to thermal effects. 

Molecular dynamics simulations were performed within a charge-neutral periodic cubic box of volume \(V = L^3\) and \(N=1024\) total electrons, with a Ewald treatment of long-range Coulomb energies and associated forces~\cite{Svensson2023}. For the low temperature condition a timestep of \(10^{-4} \, \text{fs}\) and Pauli interaction cutoff of \(12 \, a_0\) ensured energy was conserved within acceptable limits. The high temperature condition required a smaller timestep of \(5\times10^{-5}\, \text{fs}\). To collect particle configuration data for structural analysis, each simulation was initialised from a configuration that minimized the interaction energy, before a velocity scaling procedure was applied to reach the desired temperature. Specifically, both the electron and ion velocities were intermittently rescaled over a \(10 \, \text{fs}\) period to match the target temperature, followed by an additional 25 fs during which only the ion velocities were rescaled. The system was subsequently evolved for \(25 \, \text{fs}\) in the microcanonical ensemble to collect data, during which the fractional change in the total energy \(\Delta \mathcal{H}/ \mathcal{H}\) remained below \(10^{-2}\) for all simulations.

\subsection{Structural properties of the fully ionized model} \label{sec:structutral_fully_ionized}

In this subsection the role of the confining potential in the \emph{fully ionized} model is analyzed. The electronic wavefunction is therefore modeled as a product state of anisotropic Gaussians, with no bound state wavefunctions. The results for the proton-proton and electron-proton radial distribution functions (RDFs) under the two conditions and for range of different confinement widths are shown in Fig.~\ref{fig:fully_ionised_confinement_scan}. Further details on the calcultion are provided in Appendix~\ref{app:pair_correlation_functions}. First, we consider the top row, corresponding to the partially-ionized condition at \(r_s=3.23\) and \(T=55700 \, \text{K}\). A varying number of independent simulations were carried out at each confinement strength to obtain sufficient statistics, particularly to resolve weakly coupled ions at strong electron confinement. In panel (c), increasing the spatial extent \(\sigma_0\) reduces the short-range correlation in the electron-proton RDF and at \(\sigma_0 = 1.3 \, a_0\) only longer-range electron screening remains. At \(\sigma_0 = 0.5 \, a_0\), the electron-ion RDF shows a stronger short-range correlation. This peak at \(r_s \approx 0.3\) and therefore \(r \approx 1.0 \, a_0\) is a signature of bound state formation~\cite{Filinov2023}, even though no bound-state wavefunctions are present. This correlation  shoulder is also observed in the reference path integral Monte Carlo data. Two clear limits emerge: When the widths are small, \(\sigma_0 \lesssim a_0\), the potentials approach those of point particles, exhibiting deep attractive wells in the ion–electron interaction. For large widths,  \(\sigma_0 >  a_0\), the electrons are delocalized, leading to weaker interactions and reduced correlation with the ions. The corresponding ionic RDFs in panel (a) support this general picture: electronic screening determines the effective ion-ion potential, and a clear transition is observed between weakly coupled protons at \(\sigma_0 = 0.5 \, a_0\) and moderately coupled protons at \(\sigma_0 = 1.3 \, a_0\). At this condition, the structural properties are evidently sensitive to the choice of confinement strength.
\begin{figure*}
  \includegraphics[width=\linewidth]{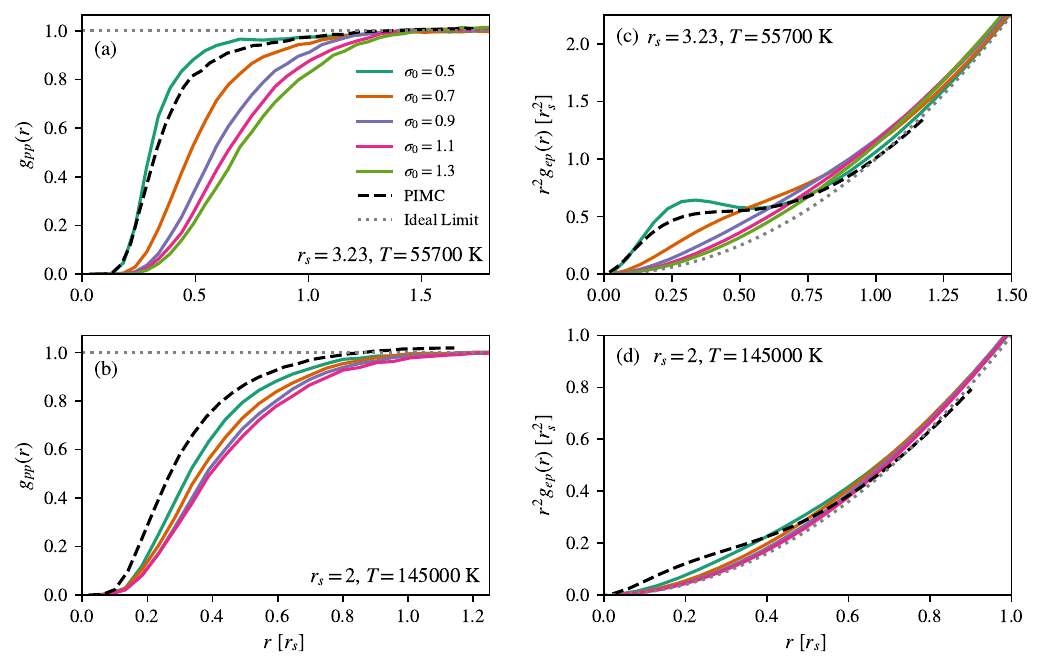}
  \caption{\label{fig:fully_ionised_confinement_scan} Radial distribution function (RDF) dependence on confinement parameter \(\sigma_0\) for the fully ionized wavepacket model. The results are compared against reference path integral Monte Carlo data from Ref.~\cite{Dornheim2024} and the ideal (uncorrelated) limit. Panels (a) and (c) correspond to the proton-proton and radially weighted electron-proton RDFs at the partially-ionized condition: \(r_s = 3.23\) and \(T=55700 \, \text{K}\). Panels (b) and (d) correspond proton-proton and radially weighted electron-proton RDFs at the fully ionized condition: \(r_s = 2\) and \(T=145000 \, \text{K}\). Hartree atomic units are used in the legend.}
\end{figure*} 
The second row in Fig. \ref{fig:fully_ionised_confinement_scan} corresponds to data for the fully ionized condition at \(r_s = 2\) and \(T = 145000 \, \text{K}\). Although a similar general trend is observed, the electron-ion RDF is greatly suppressed. This is a hallmark of weaker electron-ion coupling and therefore the resulting structural properties are less sensitive to the confinement strength, which supports previous conclusions~\cite{Svensson2024pub}. In summary, the static structure of dense plasma wavepacket simulations is most sensitive to the confining potential in the partially-ionized regime. Additionally, the electron-ion coupling -- and therefore bound state formation -- can be arbitrarily increased by confining the electrons to below the length scale of electron bound states. Regulating the spatial extent of the free electrons alters the model, and, as shown below, the resulting ionization state becomes a function of the confining parameter.

\section{Prediction of partially-ionized model} \label{sec:ionization}

\subsection{Ionization calculation}

As discussed in Sec.~\ref{sec:boundStateModel}, employing a functional form that consists of \(N_{\text{n}}\) bound electrons and \(N_{\text{e}}\) free electrons renders the equilibrium charge state distribution a quantity that must be known \emph{a priori}. In the case of hydrogen, this corresponds to evaluating the ionization state, \(\bar{z}\), which can be defined as
\begin{equation}
  \bar{z} = \frac{N_{\text{i}}}{N_{\text{n}} + N_{\text{i}}} = \frac{N_{\text{i}}}{N}
\end{equation}
for a net charge–neutral system. Recently, we introduced a free energy minimization framework to perform such a task in the context of molecular dynamics~\cite{Plummer2025}. The scheme comprises performing a set of simulations at different ionization states and computing their free energy by performing a set of thermodynamic integration calculations. This allows the resulting free energy to be numerically minimized. To perform the thermodynamic integration, all interactions, denoted by \(V\), are scaled by a coupling parameter \(\lambda\), leading to a new potential energy function
\begin{equation}
\mathcal{U}(\lambda) = \lambda V.
\end{equation} Using \(\mathcal{U}(\lambda)\) directly in molecular dynamics simulations, this definition allows the free energy per particle to be calculated with~\cite{Plummer2025,Frenkel2002}
\begin{equation}
  \frac{F}{N} = \frac{F_{\text{ideal}}}{N} + \frac{1}{N}\int_0^1 \langle V\rangle_\lambda d\lambda,
\end{equation} where \(\langle \cdot \rangle_\lambda\) is an ensemble average for a system defined by the potential energy function \(\mathcal{U}(\lambda)\) with coupling parameter \(\lambda\). To calculate the ideal free energy \(F_{\text{ideal}}\), we assume the expression for classical point particles, thereby neglecting the shape kinetic energy to make contact with the known ideal limit, as done previously~\cite{Ebeling1997}. The quantum degrees follow the classical equipartition theorem and therefore contribute additional thermal energy contributions. These additional contributions are also not included, as is common practice when extracting thermodynamic properties from WPMD simulations, see Ref.~\cite{Lavrinenko2021} for example.

\begin{figure*} 
  \includegraphics[width=\linewidth]{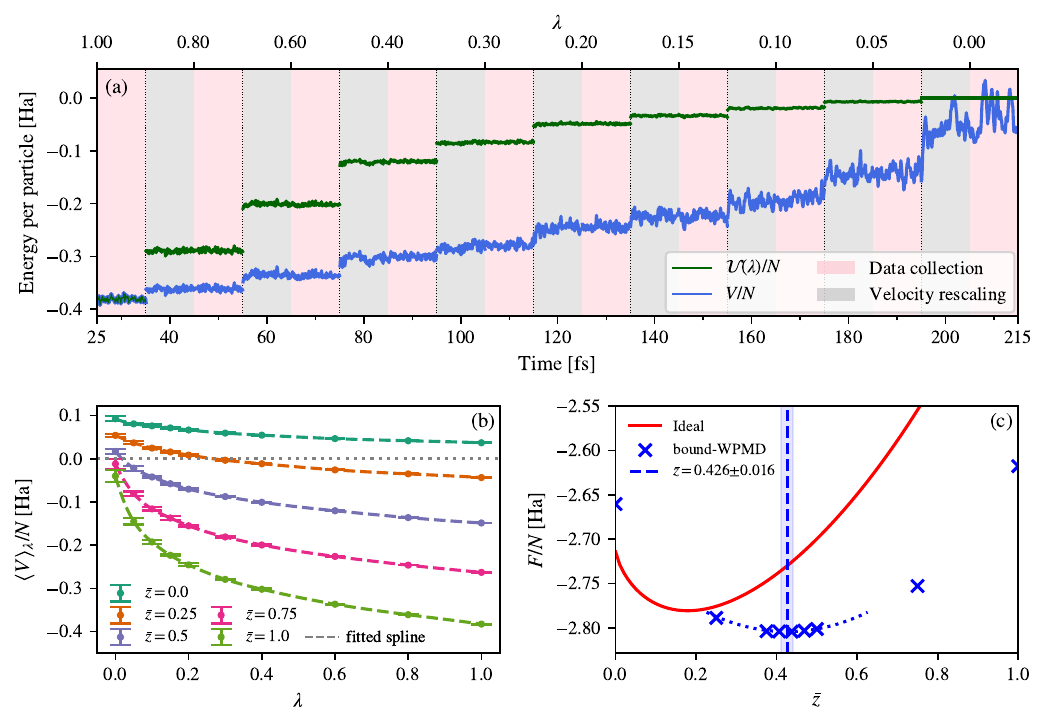}
  \caption{\label{fig:ionization_calculations}Each step of the ionization calculation for \(\sigma_0 = 1.1 \, a_0\). Panel (a) shows the time trace of the scaled potential energy function \(\mathcal{U}(\lambda)\) and the unscaled energy \(V\) across a single molecular dynamics run for \(\bar{z}=1\). The thin black dotted lines indicate the times at which the system coupling parameter is modified, with values given on the top axis. In panel (b) the resulting thermodynamic integration curves are plotted for selected ionization states \(\bar{z}\). Each data point is generated by averaging over 12 runs and taking the mean and standard deviation of the resulting ensemble averages to give the associated error bars. Panel (c) shows the resulting free energy (per particle) minimization curve generated from all sampled ionization states, including the final minima of \(\bar{z}=0.426 \pm 0.016\). The analysis method is described in more detail in Ref.~\cite{Plummer2025}.}
\end{figure*}

Molecular dynamics simulations were performed at different values of \(\lambda\) to sample the potential energy function. For a given ionization state \(\bar{z}\) and value of the coupling parameter \(\lambda\), 12 independent runs were performed for \(N=1024\) total electrons. After an initial 25 fs of velocity scaling at full coupling ($\lambda=1$), the coupling parameter was sequentially reduced in discrete steps. Each time the coupling parameter was reduced, a thermalisation step of \(10 \, \text{fs}\) was applied, in which velocity rescaling was applied to maintain the temperature of the system. Following this, data for \(\langle V\rangle_\lambda\) were collected for \(10 \, \text{fs}\), while the system was evolved in the microcanonical ensemble. This process is plotted in Fig. \ref{fig:ionization_calculations}a for one initial configuration and \(\bar{z}=1\), in which the modified potential energy \(\mathcal{U}(\lambda)\) is also shown. While the ions traverse distances approximately twice the interparticle spacing over \(10 \, \text{fs}\), averaging over multiple configurations ensures the canonical ensemble is sampled sufficiently; additionally, no large scale fluctuation of \(\mathcal{U}(\lambda)\) is present.

The results after averaging the ensemble average across 12 independent runs are plotted in Fig. \ref{fig:ionization_calculations}b for five different ionization states. The error bars represent the standard deviation associated to averaging over independent samples. Note that while these errors are small, the thermal fluctuations in the time trace may still be large, which is the case for small coupling, see Fig. \ref{fig:ionization_calculations}a. At zero coupling (\(\lambda=0\)) the full interacting potential energy function is evaluated over uncorrelated particle trajectories which is equivalent to integrating \(V\) over a uniform configuration. A positive contribution from the Pauli interactions emerges in this limit, which is repulsive. Furthermore, each particle experiences Coulomb interactions due to (1) an effective uniform background charge generated by all particles except itself, and (2) its own periodic images. Therefore, there is a negative Coulomb binding contribution at \(\lambda=0\), which is quantified and benchmarked in Appendix \ref{app:self_energy}. As the coupling parameter \(\lambda\) is increased and the corresponding interactions are scaled, particles start to interact and arrange themselves accordingly, meaning \(\langle V \rangle_{\lambda}\) decreases. As seen in our previous work~\cite{Plummer2025}, the calculation exhibits nonlinear behaviour at small coupling, followed by linear scaling around \(\lambda \approx 0.3\). To address this imbalance, low coupling values are sampled at a finer resolution. The resulting data are smoothly interpolated with a cubic spline which is integrated to find the excess free energy. Finally, these values are combined with the ideal free energy, and the result is numerically minimized using the same procedure as described in our previous work~\cite{Plummer2025}. The result for \(\sigma_0 = 1.1 \, a_0\) is shown in Fig \ref{fig:ionization_calculations}c. For \(N=1024\) total electrons, we have already shown the finite-size error is minimal when compared to the error bounds of the ionization state minimization procedure~\cite{Plummer2025}. 

This calculation was repeated for varying electron widths, with the resulting ionization states and free energies summarized in Table \ref{tab:ionization_calculations}. Numerical values for the mean electron width,
\begin{equation} 
  \langle \sigma \rangle  = \left\langle\frac{1}{N_e}\sum_{j=1}^{N_e}\frac{1}{3}\text{Tr}\left(\Sigma_j\right)\right\rangle,
\end{equation} are also presented, with the associated error from averaging over distinct runs. These results highlight that the ionization state and free energy are model-dependent, through their dependence on the confinement parameter \(\sigma_0\). As confinement strength is increased and average electronic extent reduced, the overall binding energy is increased. This reduces the effective ionization energy, because it requires less energy to promote a bound electron to a free electron. This \emph{ionization potential depression} causes an increase in the equilibrium ionization state and may be connected directly to the derivative of the excess free energy with respect to ionization state~\cite{Plummer2025}. We note that in the limit of weak confinement, the electrons will become fully delocalized and therefore comprise a uniform electron background. In this limit, the ionization state will tend to that of a mixture of neutrals and one component plasma of ions. At the considered density, the effect of neutral interactions is small and a lower bound on the ionization state can be produced by combining the ideal free energy with a one component plasma (OCP) equation of state \cite{Plummer2025}, which is also presented in Table \ref{tab:ionization_calculations} as the OCP model.
\begin{table*}[htb]
\caption{\label{tab:ionization_calculations} Ionization state $\bar{z}$, free energy $F$, and mean width \(\langle \sigma\rangle\), for different confinement parameters $\sigma_0$ in Hartree atomic units at $r_s = 3.23$ and $T = 55700 \, \text{K}$. Numbers in parentheses denote the associated uncertainty in the final digit. The error in the free energy is estimated by taking the maximum deviation between the quadratic minimum and the neighbouring free energy values adjacent to the lowest sampled data point. For comparison, the inferred path integral Monte Carlo (PIMC) ionization state from Ref.~\cite{Bellenbaum2025} for \(N=32\) atoms is also given. The one component plasma (OCP) model is defined in Ref.~\cite{Plummer2025}.}
\begin{ruledtabular}
\begin{tabular}{lclc}
$\sigma_0 \,\, [a_0]$ & $\bar{z}_{\text{min}}$ & $\langle \sigma\rangle \,\, [a_0]$ & $-F(\bar{z}_{\text{min}})/N$ [Ha] \\
\hline
$0.5 $ & $0.66 (2)$  & $0.5433(5)$ & $2.934 (3)$   \\
$0.7 $ & $0.55 (2)$  & $0.832(1)$  & $2.849 (1)$   \\
$0.9 $ & $0.47 (2)$  & $1.173(1)$  & $2.8196 (6)$  \\
$1.1 $ & $0.43 (2)$  & $1.574(2)$  & $2.804 (1)$   \\
$1.3 $ & $0.382(2)$  & $2.043(4)$  & $2.794(1)$    \\[1pt]
\hline\hline
Method & $\bar{z}_{\text{min}}$ & \multicolumn{2}{c}{Notes} \\
\hline
PIMC inferred~\cite{Bellenbaum2025} & $0.45$ & \multicolumn{2}{c}{Chemical model fit to imaginary-time density–density correlation function} \\
OCP model~\cite{Plummer2025}       & $0.24$ & \multicolumn{2}{c}{Scalar minimization} \\
Ideal                              & $0.18$ & \multicolumn{2}{c}{Scalar minimization} \\
\end{tabular}
\end{ruledtabular}
\end{table*}

\begin{figure}
  \includegraphics[width=\linewidth]{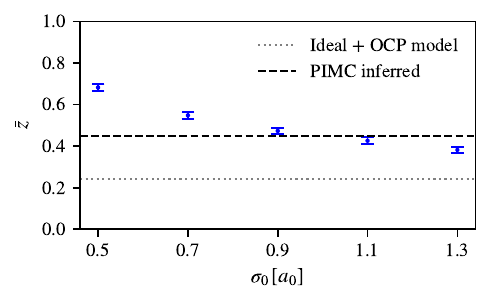}
  \caption{\label{fig:ionization_versus_confining} Minimized ionization state \(\bar{z}_{\text{min}}\) plotted against confinement parameter \(\sigma_0\) at $r_s = 3.23$ and $T = 55700 \, \text{K}$. Results are presented alongside path integral Monte Carlo inferred results from Ref \cite{Bellenbaum2025} and the one component plasma (OCP) model from Ref. \cite{Plummer2025}. This data may be found in Table~\ref{tab:ionization_calculations}.}
\end{figure}

Ionization states may be inferred from path integral Monte Carlo simulations in a variety of ways, see~\cite{Bellenbaum2025,bonitz2024pub,Filinov2023,Bonitz2025, Dornheim2025}, although results can differ between approaches. Direct comparison is possible with Ref.~\cite{Bellenbaum2025}, where a chemical model employing the Chihara decomposition was fitted to the imaginary-time density–density correlation function; in Fig.~\ref{fig:ionization_versus_confining}, relatively close agreement is found for confinement widths \(\sigma_0 = 0.9 \, a_0\) and \(\sigma_0 = 1.1 \, a_0\). We also note that a trajectory analysis approach~\cite{Filinov2023} could be used to determined the number of free electrons (anisotropic wavepackets) that are effectively bound in the simulation. Such an approach may reduce the overall sensitivity in the final ionization state. However, given that the ionization state is definition-dependent, a rigorous comparison based on structural properties is preferred, and performed in the following section.

\subsection{Impact on Structural Properties}
\begin{figure*}
  \includegraphics[width=\linewidth]{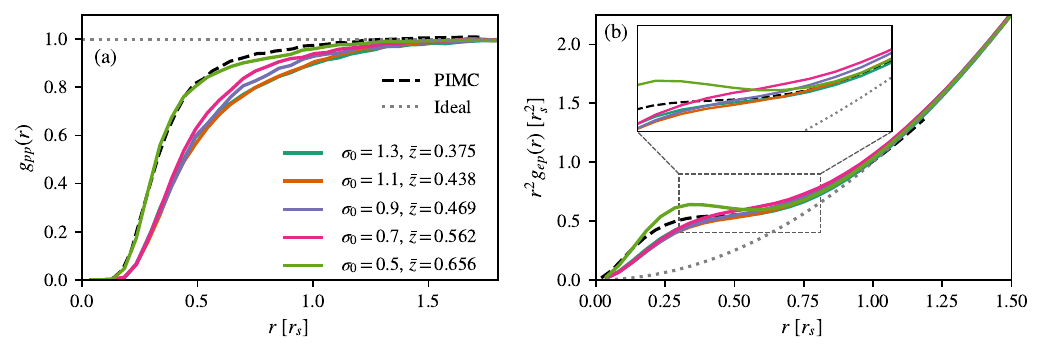}
  \caption{Radial distribution functions (RDFs) from the bound-WPMD model for different values of the confinement parameter \(\sigma_0\). Self-consistent ionization states are computed via free energy minimization and found in Table~\ref{tab:ionization_calculations}. (a) proton-proton (pp). (b) electron-proton (ep).}\label{fig:partially_ionised_rdfs_confinement_scan}
\end{figure*}

Fig.~\ref{fig:partially_ionised_rdfs_confinement_scan} shows the structural properties of the partially-ionized model, henceforth referred to as the bound-WPMD model, evaluated at the computed equilibrium ionization states. The method by which the bound electron RDFs are calculated is discussed below. In Fig.~\ref{fig:partially_ionised_rdfs_confinement_scan}b, the inclusion of bound-state wavefunctions reduces the sensitivity of the electron-proton correlations for \(\sigma_0\geq0.7 \, a_0\) when compared with the fully ionized model. At these conditions the electron screening agrees moderately well with the path integral Monte Carlo reference, although the correlation shoulder is slightly diminished at small separation. At \(\sigma_0=0.5 \, a_0\), electron–proton correlations are pronounced, comparable to the fully ionized result shown in Fig.~\ref{fig:fully_ionised_confinement_scan}c. The associated ion structure in Fig.~\ref{fig:partially_ionised_rdfs_confinement_scan}a informs a similar picture. Surprisingly, the enhanced electron-ion interaction for narrow electron widths yields the closest ion structure to the reference data.
As discussed in Sec. \ref{sec:structutral_fully_ionized}, imposing a confinement length scale for the free electrons that is smaller than the characteristic size of the bound states artificially enhances the electron–proton interaction strength. This scenario is unphysical, as genuinely free electrons are expected to exhibit only weak attraction to the ionic cores. This is supported by the very strong electron-ion correlation observed in both the partially and fully ionized results. Furthermore, although ionization state has a degree of choice in its definition, best agreement is reached for intermediate confinement parameters in Fig.~\ref{fig:ionization_versus_confining}. Therefore, we proceed with an in-depth analysis of \(\sigma_0 = 1.1 \, a_0\), which corresponds to an interacting mean width of \(\langle \sigma \rangle = 1.57 \, a_0\). Returning to Fig. \ref{fig:fully_ionised_confinement_scan}c, this confinement strength provides some electron-proton long-range screening as would be expected in a fully-ionized system while the electrons are not confined below the extent of the 1s wavefunctions employed within the bound state model.
\begin{figure*}
  \includegraphics[width=\linewidth]{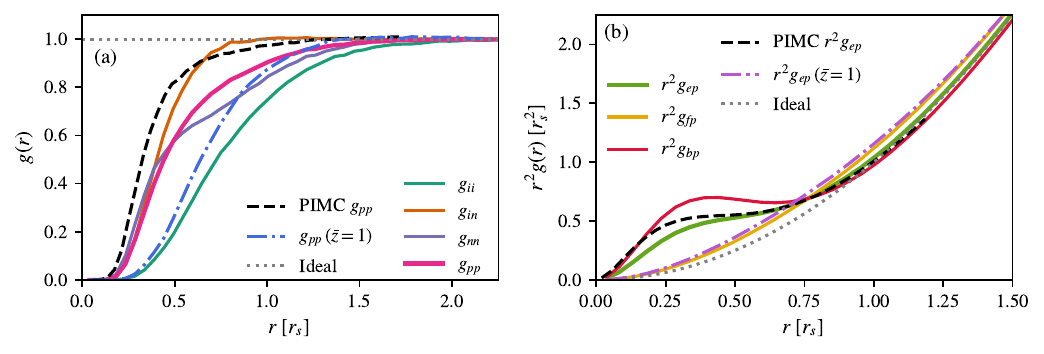}
  \caption{\label{fig:partially_ionised_rdfs}Radial distribution functions (RDFs) at $r_s = 3.23$ and $T = 55{,}700\ \mathrm{K}$ computed with the bound-WPMD model for $\bar{z} = 0.44$ and $\sigma_0 = 1.1 \, a_0$, compared with path-integral Monte Carlo data, fully ionized ($\bar{z} = 1$) results, and the ideal uncorrelated case. (a) Ion-ion (${ii}$), ion--neutral (${in}$), and neutral--neutral (${nn}$) RDFs, combined via Eq.~\eqref{eq:RDF_pp_decomp} to yield the total proton-proton RDF ($pp$). (b) Free-proton ($fp$) and bound-proton (${bp}$) RDFs, combined via Eq.~\eqref{eq:RDF_ep_decomp} to yield the total electron-proton RDF ($ep$).}
\end{figure*}

\begin{figure}
  \includegraphics[width=\linewidth]{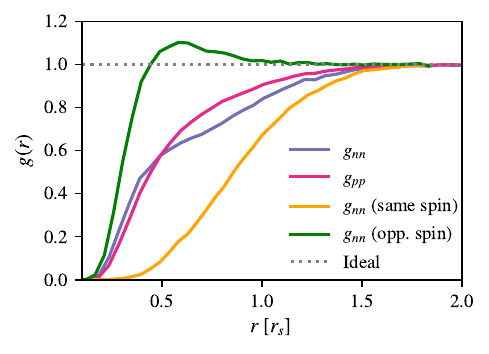}
  \caption{\label{fig:spin_resolved_rdfs}Spin-resolved neutral-neutral radial distribution functions (RDFs) at $r_s = 3.23$ and $T = 55{,}700\ \mathrm{K}$ computed with the bound-WPMD model for $\bar{z} = 0.44$ and $\sigma_0 =1.1 \, a_0$.
  For comparison, the total neutral-neutral and proton-proton RDFs in Fig.~\ref{fig:partially_ionised_rdfs} are also plotted.}
\end{figure}

The full electron-proton RDF is calculated by decomposing into bound and free contributions:
\begin{equation} \label{eq:RDF_ep_decomp}
  g_{ep} = \frac{\bar{z}^2g_{pf} + (1-\bar{z})g_{pb}}{\bar{z}^2-\bar{z} + 1},
\end{equation}
which is exhibited in Fig.~\ref{fig:partially_ionised_rdfs}. Here the proton-free and proton-bound RDFs, \(g_{pf}\) and \(g_{pb}\) respectively, are computed through the sampling algorithm presented in Appendix \ref{app:pair_correlation_functions}. The proton-bound RDF exhibits a clear signature of the bound electrons present in the model, while the proton-free RDF shows longer range screening as observed in the results for \(\bar{z}=1\). 

Similarly to Eq. \eqref{eq:RDF_ep_decomp}, a decomposition of the proton-proton RDF
\begin{equation} \label{eq:RDF_pp_decomp}
  g_{pp} = \bar{z}^2 g_{ii} + 2\bar{z}(1-\bar{z})g_{in} + (1-\bar{z})^2g_{nn}
\end{equation}
into ion-ion, ion-neutral and neutral-neutral RDFs denoted by \(g_{ii}\), \(g_{in}\) and \(g_{nn}\) respectively, is possible. Inspecting Fig. \ref{fig:partially_ionised_rdfs}a, the ion-ion correlation remains moderately coupled. In fact, plotting the ion-ion RDF \(g_{ii}\) against an effective Wigner-Seitz radius \(\bar{z}^{1/3} r_s\) exposes the ion coupling to be slightly reduced when compared with the proton-proton RDF \(g_{pp}\) at \(\bar{z}=1\). The neutral-neutral and ion-neutral RDFs both show weaker short-range correlation, as expected given their heavily screened Coulomb interactions and the repulsive Pauli interactions between same-spin bound electrons. As already discussed, the final combined proton-proton RDF is in better agreement with the path integral Monte Carlo data when compared with the fully-ionized model, although some discrepancy remains. Although the reference data inherently contain finite-size errors, it is nevertheless instructive to consider possible sources of model error within the bound-WPMD model. One explanation could be due to slightly reduced screening observed in the electron-proton correlations, although we note that a similar effect is seen for the ion-ion RDFs in the fully ionized case in Fig. \ref{fig:fully_ionised_confinement_scan}b. An alternative explanation is that omission of bound-state polarization artificially enhances the apparent proton–proton coupling in the RDFs; including polarization effects would possibly allow transient molecular-ion ($H_2^{+}$) formation and thus reduce the effective coupling. In contrast, the opposite spin neutral-neutral RDF, in Fig.~\ref{fig:spin_resolved_rdfs}, exhibits a peak indicative of liquid-like structure. This peak originates from the attractive asymptote in the neutral-neutral Coulomb potential, Eq.~\eqref{eq:neutral_neutral_coulomb}.

\section{Concluding Remarks} \label{sec:conclusion}

We have introduced the bound-WPMD model with an explicit representation of bound electrons, hence requiring a self-consistent ionization calculation. This was carried out using the framework developed in Ref.~\cite{Plummer2025}, based on thermodynamic integration and free energy minimization. While earlier works have modeled bound-electron effects using semi-empirical effective core potentials in the context of WDM and dynamic electron chemistry \cite{Davis2020, Xiao2015}, explicit charge-state distributions have not been considered previously.

Applying this method to dense, partially-ionized hydrogen has enabled a direct comparison of the resulting structural properties with reference path-integral Monte Carlo data.  
As demonstrated in Sec. \ref{sec:structutral_fully_ionized}, the confining potential regulates the free electron-ion interaction within the model. Therefore, at physical conditions where strong electron-ion coupling is expected i.e. partially-ionized conditions, the resulting structural properties are particularly sensitive to its choice. Introducing the partially-ionized model reduces this sensitivity. For weak confinement there is less electron-ion interaction which reduces the ionization potential depression. This results in a smaller ionization state and therefore increases the electron-ion correlation through inclusion of more bound electrons. For stronger free electron confinement there is greater ionization potential depression, which reduces the number of bound electrons. However, the free electrons have smaller spatial extents and bind more strongly to the ions. When imposing physically motivated constraints on the free electron extent, the structural properties of the bound-WPMD model are in better agreement with the selected reference data. This assessment is only possible with a systematic method to find the ionization state.

Generally the inclusion of bound electrons could be used as a scheme in other empirical dynamic electron molecular dynamics models for warm dense matter, where weaker-than-expected electron-ion coupling has been identified, for example Ref.~\cite{Campbell2025}. We emphasize that dense, partially-ionized plasma is a particularly challenging regime to model; in fact, high-fidelity reference data for structural properties have only recently become available~\cite{bonitz2024pub,Dornheim2024} and extension of molecular dynamics methods into this regime is ongoing~\cite{Demyanov2025,Lavrinenko2021}. Our model could be extended to molecular hydrogen through the appropriate selection of Pauli potentials~\cite{Angermeier2021}, or by introducing molecules and molecular ions as additional species and performing additional free energy minimizations. Such an approach would increase the dimensions of the minimization, although it may be possible to find systematic ways to streamline the free energy calculation~\cite{Frenkel2002}. Additionally, the description of higher atomic number, non-bonded and partially-ionized dense plasma is possible within the framework. This extension would require the consideration of bound states with higher quantum number.

Finally, we note that it may be possible to include stochastic transitions between bound and free state manifolds systematically within the framework of wavepacket models~\cite{Feldmeier2000, Ono2004}, providing an interesting course of future research for nonequilibrium modeling of plasmas using dynamic, restricted wavefunction approaches.

\begin{acknowledgments}
  We gratefully acknowledge helpful comments from Tobias Dornheim regarding the path integral Monte Carlo method. We are also grateful for computing resources provided by STFC Scientific Computing Department’s SCARF cluster. DP, PS, SMV and GG acknowledge support from AWE-NST UK via Oxford Centre for High Energy Density Science (OxCHEDs). PS acknowledges funding from the Oxford Physics Endowment for Graduates (OXPEG). S.M.V. acknowledges support from the UK EPSRC grant EP/W010097/1. The work of SMV and GG has received partial support from EPSRC and First Light Fusion under the AMPLIFI Prosperity partnership, grant no. EP/X025 373/1. The work by PS was partially supported by the Center for Advanced Systems Understanding (CASUS), financed by Germany’s Federal Ministry of Education and Research and the Saxon state government out of the State budget approved by the Saxon State Parliament, and the European Research Council (ERC) under the European Union’s Horizon 2022 research and innovation programme (Grant agreement No. 101076233, ``PREXTREME”). Views and opinions expressed are however those of the authors only and do not necessarily reflect those of the European Union or the European Research Council Executive Agency. Neither the European Union nor the granting authority can be held responsible for them.
\end{acknowledgments}

\section*{Data Availability}
  The data underlying the figures in the main text are available in Ref.~\cite{figure_data}.

\appendix

\section{Gaussian decomposition}
\label{app:gauss_decomp}

\begin{figure}
    \includegraphics[width=\linewidth]{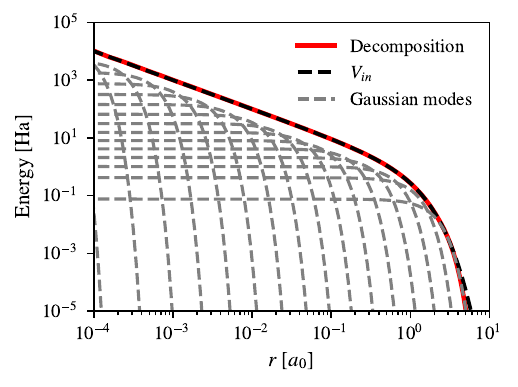}
  \caption{\label{fig:gauss_decomp} Decomposition of the ion-neutral kernel used internally within the \textsc{LAMMPS} implementation to compute the electron-neutral Coulomb energy. Individual Gaussian modes are depicted with grey dashed lines and the sum over these modes is shown in red and closely match the target (black dashed) line.}
\end{figure}

The \(c_p\) and \(\alpha_p\) coefficients in Eq. \eqref{eq:ion_neutral_gauss_decomp} are obtained using the method presented in Ref.~\cite{Svensson2023}, which is elaborated here. The integrated squared error weighted by the radial shell volume,
\begin{equation}
    L = \int_0^{r_{max}} dr \, r^2 \left[V(r) - \sum_p c_p \text{e}^{-\alpha_p r^2}\right]^2,
\end{equation}
was minimized to find a Gaussian decomposition of \(V(r)\). To obtain an analytical formula for the amplitudes, the derivative of the loss function with respect to \(c_p\) may be taken, resulting in
\begin{equation}
    \frac{\partial L}{\partial c_p} = \int dr \, 2r^2 \left[\sum_{p'} c_{p'} \text{e}^{-(\alpha_p + \alpha_{p'})r^2} - V(r) \text{e}^{-\alpha_p r^2}\right].
\end{equation}
Employing the minimization condition \(0 = {\partial L}/{\partial c_p}\)
and using known formulae for Gaussian integrals gives the following linear equation, 
\begin{equation}
  \boldsymbol{b} = \text{A} \boldsymbol{c},
\end{equation}
under the following identifications:
\begin{align}
    [\boldsymbol{c}]_p &= c_p \\ \qquad [A]_{pq} &= \frac{1}{4}\left(\frac{\pi}{(\alpha_p + \alpha_{q})^3}\right)^{1/2} \\ [\boldsymbol{b}]_p &= \int dr \, r^2V(r) \text{e}^{-\alpha_p r^2}.
\end{align}
The optimal \(\{c_p\}\) coefficients may therefore be obtained through inversion of the \(\text{A}\) matrix. Therefore, the loss function was numerically minimized with respect to the width coefficients \(\{\alpha_p\}\) preconditioned with the optimal amplitude coefficients \(\{c_p\}\). With this new notation, the loss is expressed as
\begin{equation}
\begin{split}\label{eq:loss_matrix_expr}
    L(\{\alpha_p\}) &= \boldsymbol{c}^\text{T} \text{A} \boldsymbol{c} - 2\boldsymbol{c}^\text{T} \boldsymbol{b} + \int dr \, r^2 V^2(r).
\end{split}
\end{equation}
For the case of the ion-neutral interaction, it is simple to find an explicit form for the \(\boldsymbol{b}\) integral and the final constant term in Eq. \eqref{eq:loss_matrix_expr}. To achieve a stable numerical minimization algorithm, \(\alpha_p\) modes were added sequentially. The ion-neutral decomposition used for the simulations performed here is given in Fig. \ref{fig:gauss_decomp}. This decomposition contains 18 modes with a loss of \(L < 2\times10^{-7}\) in Hartree atomic units.

\section{Expression for electron-neutral Coulomb interaction} \label{app:electron_neutral_coulomb}
To calculate the Coulomb energy through a Gaussian decomposition, the sum over the following integral is required:
\begin{equation}
  I_p = \int d^3\boldsymbol{x}|\langle \boldsymbol{x}|q_i \rangle |^2 \exp\left[-\alpha_p(\boldsymbol{x} - \boldsymbol{r}_j)^2\right]
\end{equation}
where \(p\) is the index of the respective mode. This integral may be evaluated through standard multidimensional Gaussian integral formulae. A similar expression is found in Appendix A of Ref.~\cite{Ono2012}. Performing the integral and applying the following identity, which is a special case of Eq. (24) in Ref.~\cite{Henderson1981},
\begin{equation*} \label{eq:unsimplified_matrix}
    \Sigma^{-1}_i - \Sigma^{-1}_i \left(\Sigma^{-1}_i + 2\alpha_p\text{I}\right)^{-1} \Sigma^{-1}_i = 2\alpha_p \left[\text{I} + 2\alpha_p \Sigma_i\right]^{-1},
\end{equation*}
where \(\text{I}\) is the identity matrix, yields the following formula
\begin{equation}
     I_p = \frac{\exp\left[-\boldsymbol{r}^{\text{T}}\left(\alpha_p \left[\text{I} + 2\alpha_p \Sigma_i\right]^{-1}\right)\boldsymbol{r}\right]}{\sqrt{\det\left(\text{I} + 2\alpha_p\Sigma_i\right)}}.
\end{equation}
Here \(\boldsymbol{r} = \boldsymbol{r}_i - \boldsymbol{r}_j\). This is the final expression used within the mode sum to compute the Coulomb interactions of the model. For example, the electron-neutral pair interaction is
\begin{equation}
  \mathcal{V}_{en}^C(\boldsymbol{r}_n) = \sum_{p} c_p I_p.
\end{equation}
The internal \textsc{LAMMPS} implementation sums from large to small values of \(\alpha_p\), truncating the sum when energy contributions are negligible. This is motivated by the fact that some modes are only relevant at small interparticle separations, see Fig. \ref{fig:gauss_decomp}.

\section{Matrix elements for Pauli interactions} \label{app:kinetic_matrix_elements}

To compute the Pauli interaction on-the-fly, explicit formulae are required for the overlap and kinetic energy matrix elements between anisotropic Gaussians \(|q_i\rangle\) and isotropic Gaussians \(|g_{j,p}\rangle\). 
To write down concise expressions for the required quantities the following notatation is introduced:
\begin{align}
    \label{eq:kij}\boldsymbol{k}_{ij} &= \frac{\boldsymbol{p}_i - \boldsymbol{p}_j}{\hbar} \\
    a_{ij} &= \boldsymbol{r}_i^{\rm{T}} {\rm{M}}_i \boldsymbol{r}_i + \boldsymbol{r}_j^{\rm{T}}{\rm{M}}_j^* \boldsymbol{r}_j. \\
    \boldsymbol{\Delta}_{ij} &= {\rm{M}}_{ij}^{-1} \boldsymbol{c}_{ij} \\
    \boldsymbol{c}_{ij} &= {\rm{M}}_i \boldsymbol{r}_i  + {\rm{M}}_j^* \boldsymbol{r}_j \\
    {\rm{M}}_{ij} & = {\rm{M}}_i + {\rm{M}}_j^* \\
    \label{eq:Mi}{\rm{M}}_i & = \frac{1}{4}\Sigma_i^{-1} - \frac{i}{\hbar} \Pi_i.
\end{align}
For an isotropic Gaussian \(|g_{j,p}\rangle\), as defined in Eq. \eqref{eq:propagating_isotropic_gaussian},
the internal degrees of freedom reduce to
\begin{equation}
  \Sigma_j = \frac{1}{4\alpha_{p}} \text{I}\quad ; \quad \Pi_j = 0.
\end{equation}
Therefore, the general expressions for anisotropic matrix elements are applicable using the notation in equations \eqref{eq:kij}--\eqref{eq:Mi} with minimal alteration. The overlap between two anisotropic wavepackets follows from known formulae for multidimensional Gaussian integrals:
\begin{equation}    
  \begin{split}
    \langle q_i | q_j \rangle = &\left(8\det\left(\text{M}^*_{ij}\right)\sqrt{\det\left(\Sigma_i\right)\det\left(\Sigma_j\right)}\right)^{-1/2} \\ & \times \exp\bigg[(\boldsymbol{\Delta}_{ij}^*)^{\text{T}}\boldsymbol{c}_{ij}^* - i(\boldsymbol{\Delta}_{ij}^*)^{\text{T}}\boldsymbol{k}_{ij} \\ & -\frac{1}{4}\boldsymbol{k}_{ij}^{\text{T}}(\text{M}_{ij}^*)^{-1}\boldsymbol{k}_{ij} - a_{ij}^* + \frac{i}{\hbar}(\boldsymbol{p}_i^{\text{T}}\boldsymbol{r}_i - \boldsymbol{p}_j^{\text{T}}\boldsymbol{r}_j)\bigg].
  \end{split}
\end{equation}
The kinetic energy matrix element may be calculated by equating moments of the product of the real-space wavefunctions to derivatives of the overlap integral, i.e. differentiation under the integral sign. Doing the calculation yields the following result for the mean displacement:
\begin{equation}
    \boldsymbol{d}_{ij} = \frac{\langle q_i | \hat{\boldsymbol{x}} - \boldsymbol{r}_j | q_j \rangle }{\langle q_i | q_j \rangle} = -\frac{i}{2}\left(\text{M}_{ij}^*\right)^{-1}\boldsymbol{k}_{ij} + \boldsymbol{\Delta}_{ij}^* - \boldsymbol{r}_j.
\end{equation}
which corresponds to the first moment vector. The second-order moment tensor is
\begin{equation}
  \begin{split}
    \text{N}_{ij} & = \frac{\langle q_i | (\hat{\boldsymbol{x}} - \boldsymbol{r}_j)(\hat{\boldsymbol{x}} - \boldsymbol{r}_j)^{\text{T}}|q_j\rangle}{\langle q_i | q_j \rangle} \\ & = \frac{1}{2}\left(\text{M}_{ij}^*\right)^{-1} + \boldsymbol{d}_{ij} \boldsymbol{d}_{ij}^{\text{T}}.
  \end{split}
\end{equation} 
With these results, the kinetic energy may now be written concisely:
\begin{equation}
  \begin{split}
    \langle q_i | \frac{\boldsymbol{p}^2}{2m} | q_j \rangle = \bigg[ & \frac{\hbar^2}{m}+ \text{Tr}\left\{\text{M}_j\right\} \frac{\boldsymbol{p}_j^2}{2m} +  \frac{2i\hbar}{m} \boldsymbol{p}_j^{\text{T}} \text{M}_j \boldsymbol{d}_{ij} \\ &- \frac{2\hbar^2}{m} \text{Tr}(\text{M}_j \text{M}_j \text{N}_{ij})\bigg] \langle q_i | q_j \rangle.
  \end{split}
\end{equation}
Using these formulae, the free-free kinetic Pauli potential may be derived, an explicit formula is given in Ref.~\cite{Svensson2024thesis}. For the free-bound interaction, the relevant matrix elements are computed with the following propagating Gaussian mode sums:
\begin{align}
  \langle q_j| b_k \rangle &= \sum_{p} c_p \langle q_j | g_{k,p} \rangle \\
  \langle q_j| \hat{\mathbf{p}}^2| b_k \rangle&= \sum_{p} c_p \langle q_j | \hat{\mathbf{p}}^2 | g_{k,p} \rangle.
\end{align}
Where each Gaussian wavefunction is given in Eq. \eqref{eq:propagating_isotropic_gaussian}. Application of Eq. \eqref{eq:pauli_expanded} gives the free-bound kinetic Pauli potential. To compute the forces, analytical expressions for the matrix element derivatives are required.

\section{Calculation of radial distribution functions}
\label{app:pair_correlation_functions}
The general formula for the instantaneous radial distribution function between a species' \(a\) and \(b\) with respective indices \(a_i\) and \(b_j\) is~\cite{Ono2012}
\begin{equation} \label{eq:RDF_general}
\begin{split}
  g_{ab}(\boldsymbol{x}) &= \frac{V}{N_a (N_b - \delta_{ab})}\left\langle \sum_{a_i \neq b_j} \delta^3\left(\boldsymbol{x} - [\hat{\boldsymbol{x}}_{a_i} - \hat{\boldsymbol{x}}_{b_j} ]\right)\right\rangle
\end{split}
\end{equation}
where  \(N_a\) is the number of particles corresponding to species \(a\), \(\delta_{ab}\) is the kronecker delta function which appears to correctly normalize the self RDF, and \(\delta^3(\boldsymbol{x})\) is the 3 dimensional Dirac delta function. Within the context of isotropic systems, the RDF is typically ensemble averaged and further averaged over spherical shells of width \(\delta r\) to give \(\bar{g}(r)\). For point particles (the ions), it is possible to express the RDF as a histogram,
\begin{equation}
\begin{split}\label{eq:point_particle_RDF}
  \bar{g}_{ab}(r) = & \frac{1}{V_{\text{shell}}(r, \delta r)}\frac{V}{N_a (N_b - \delta_{ab})}\\ & \times \sum_{a_i \neq b_j} \mathbb{I}(r \le |\boldsymbol{r}_{a_i} - \boldsymbol{r}_{b_j}| < r + \delta r)
\end{split}
\end{equation}
where \(\mathbb{I}\) is an indicator function that returns one if the separation lies within the spherical shell defined by the interval \([r, r+\delta r)\), and zero otherwise. To compute the averaged RDF for arbitrary wavefunctions, we employ the following generalization of Eq. \eqref{eq:point_particle_RDF}:
\begin{equation}
\begin{split}\label{eq:wavefunction_RDF}
  \bar{g}_{ab}(r) = & \frac{1}{V_{\text{shell}}(r, \delta r)}\frac{V}{N_a (N_b - \delta_{ab})}\\ & \times \frac{1}{N_s}\sum_{k=1}^{N_s}\sum_{a_i \neq b_j} \mathbb{I}(r \le |\boldsymbol{y}^{k}_{a_i b_j}| < r + \delta r).
\end{split}
\end{equation}
Here the simple binning of pairwise distances has been replaced with \(N_s\) independent samples, denoted by \(\boldsymbol{y}^{k}_{a_i b_j}\). For proton-free electron RDFs, these samples are generated from the following probability distribution:
\begin{equation}
\begin{split}
P_{i k}(\boldsymbol{y})
  =&
   \langle k |
        \delta^3\left(\mathbf{y} - [\hat{\mathbf{x}}_{k}-\boldsymbol{r}_{i}]\right)
      |k \rangle \\
      = &\frac{\exp\left[-\frac{1}{2}(\boldsymbol{y}-\boldsymbol{r}_{k i})^{\text{T}}\Sigma_{k}^{-1}(\boldsymbol{y}-\boldsymbol{r}_{k i})\right]}{\left((2\pi)^3 \det{\Sigma_{k}}\right)^{1/2}}.
\label{eq:p_y_def}
\end{split}
\end{equation}
Similarly, for proton-bound calculations, the corresponding probability distribution is
\begin{equation}
\begin{split}
  P_{i k}(\boldsymbol{y})
  &= \langle b_{k} | \delta^3\left(\mathbf{y} - [\hat{\mathbf{x}}_{k}-\boldsymbol{r}_{i}] \right)|b_{k}\rangle \\
  &= \frac{1}{a_0^3 \pi} \times \exp\left[- \frac{2|\boldsymbol{y} - \boldsymbol{r}_{i k}|}{a_0}\right].
\end{split}
\end{equation}
Changing coordinates to \(\boldsymbol{z} = \boldsymbol{y} -\boldsymbol{r}_{i k}\) and radially integrating results in the Gamma function with shape parameter \(k=3\) and scale parameter \(\theta=\frac{1}{2}\). This enables a fast sampling routine, given that the Gamma distribution can be directly sampled with common numerical libraries. Once radial distances are sampled, angular coordinates are generated from a uniform distribution over the unit sphere to find \(\boldsymbol{z}\). These are then linearly translated to find samples for \(\boldsymbol{y} = \boldsymbol{z} + \boldsymbol{r}_{i k}\).

\section{Self energy Coulomb contributions at zero coupling}
\label{app:self_energy}

\begin{figure}
    \vspace{0.1cm}
    \includegraphics[width=\linewidth]{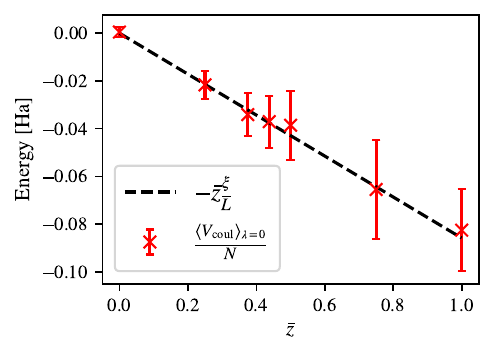}
  \caption{\label{fig:ave_coulomb_zero_coupling} Ensemble average of Coulomb energy at zero coupling, compared against the Madelung energy. Here \(\sigma_0 = 1.1\) and the error bars are generated in the same manner as Fig. \ref{fig:ionization_calculations}b.}
\end{figure}

At zero coupling all particle trajectories are uncorrelated. Therefore, a given particle (ion or Gaussian wavepacket), will experience a uniform charge density that neutralizes its own charge. Additionally, it will interact with its own periodic replicas. Therefore the self-energy at zero coupling is equal to the Madelung energy~\cite{Slattery1980}, which, assuming the spatial extent of the electrons is considerably smaller than the box length \(L\) is given by
\begin{equation} \label{eq:madelung_term}
  \langle V_{\text{coul}} \rangle_{\lambda=0} = \frac{e^2}{4\pi\epsilon_0}\frac{\xi}{2L} 2N.
\end{equation} Here \(V_{\text{coul}}\) is the Coulomb interactions of the model and \(\xi = 2.837297479\)~\cite{Slattery1980} is the madelung constant of a simple cubic lattice of ions with unit spacing. The short-range neutral Coulomb interactions average to zero when \(L\gg a_0\)~\cite{Plummer2025}. To benchmark the thermodynamic dynamic integration calculation, \(\langle V_{\text{coul}} \rangle_{\lambda=0}\) was explicitly calculated and is plotted across ionization state in Fig. \ref{fig:ave_coulomb_zero_coupling}. As can be seen, Eq. \eqref{eq:madelung_term} holds within statistical error.

\bibliography{library}

\end{document}